\documentclass[pre,amsmath]{revtex4-2}
\usepackage{graphicx}
\usepackage{color}
\usepackage{bm}
\usepackage{epsfig}
\usepackage{latexsym}
\usepackage{nameref,hyperref}
\usepackage{pifont}
\usepackage[utf8]{inputenc}
\usepackage{subcaption}
\usepackage{csquotes}
\usepackage{enumitem}
\usepackage{bm}
\usepackage{float}
\usepackage{xcolor}
\usepackage{multirow}
\usepackage{amsmath}
\usepackage{amssymb}
\usepackage{booktabs}
\usepackage[utf8]{inputenc}
\usepackage[T1]{fontenc}
\usepackage{lmodern} 
\usepackage{booktabs}

\DeclareUnicodeCharacter{0229}{\k{e}}

\begin{document}
\title{Phase ordering kinetics in Light-Heavy-Vacancy model: unusual coarsening dynamics}
\author{Chandradip Khamrai and Sakuntala Chatterjee}
\affiliation{Department of Physics of Complex Systems, S. N. Bose National Centre for Basic Sciences, Block JD, Sector 3, Salt Lake, Kolkata 700106, India.}

\begin{abstract}
We study a one dimensional lattice model of coupled driven system where two kinds of hardcore particle species, `light' and `heavy' move on a fluctuating landscape. Light particles prefer to move upward along the local height gradient of the landscape, and heavy particles prefer to move downhill. In addition, these particle species also exert bias on the local height profile of the landscape in the upward or downward direction. The unoccupied or vacant parts of the landscape experience no bias and undergo symmetric height fluctuations. In an earlier work, we had derived a phase diagram of the system which consists of different kinds of ordered and disordered phases. In an ordered phase one or both particle species undergo phase separation and the landscape can also organize itself in the form of a large hill or deep valley. In this work we study the coarsening properties where starting from a disordered state we monitor how long range order develops for particles and landscape as time goes on. Unlike conventional phase ordering systems, where coarsening process progresses by formation of small ordered domains and subsequent merger of those domains, here we find ordered domains which develop at early times become unsustainable later on. Instead of merging together, these early domains disintegrate and new domains emerge which finally give rise to large scale ordered structure in the long time limit. This results in highly unusual coarsening behavior in the system. For particle coarsening we find the characteristic length scale grows with two distinctly different power law exponents at early and late times. The landscape coarsening is even more dramatic where the length scale does not even increase continuously with time, but decreases for brief time-intervals. The height fluctuations of the landscape shows periodic oscillations with time during the coarsening phase. We explain using linear hydrodynamics that this is caused by three normal modes which  move through the system like traveling waves. We calculate the propagation velocities of these modes within mean field approximation.

\end{abstract}
\maketitle

\section{Introduction}







Phase ordering kinetics \cite{bray2002theory, allen1979microscopic, puri2009kinetics} provides a general framework for understanding how long range order sets in as a system undergoes time-evolution from an initially disordered state to an ordered one. A central feature of this dynamics is coarsening, during which domains of the emerging ordered phase grow in size. Coarsening dynamics is often characterized by a time-dependent length scale that represents typical size of an ordered domain at a give time \cite{corberi2011growing, cugliandolo2015coarsening, cugliandolo2017geometric}. As ordered domains grow larger in size, this length scale also increases with time. In many phase ordering systems this increase happens in the form of a power law with an universal exponent that is determined by relevant conservation laws or symmetries {\sl etc.} present in the system. For example, depending on whether the total magnetization is conserved or not, a spin system shows coarsening with an exponent $1/3$ or $1/2$, respectively \cite{bray1994growth, onuki2002phase, krapivsky2010kinetic}. Certain phase ordering systems also show non-algebraic coarsening \cite{huse1987critical, yurke1992coarsening}. For example, while relaxing towards the ordered state, a system can get stuck in metastable states  leading to much slower, logarithmic growth of the coarsening length scale \cite{evans1998phase, Evans1998phase, chakrabortylarge, lahiri1997steadily, lahiri2000strong}.

Despite these diverse growth laws, coarsening process is always characterized by a length scale that grows monotonically with time, reflecting the progressive elimination of small-scale structures and the coalescence of larger domains. In this work, we study coarsening in a coupled driven system where the characteristic length scale shows non-monotonicity with time. Starting from a disordered state, as the system develops order, a growing length scale does emerge, whose value shows an overall increase with time. But this increase is not monotonic and for intermediate times the length scale also shows brief periods of decline. We argue that the presence of kinematic waves during the coarsening phase results in this unusual behavior.

Coupled nonequilibrium systems \cite{bisht2019interface, sediments, ramaswamy2002phase, passiveslider1, passiveslider2, singleactiveslider} can exhibit rich behavior ranging from ordered \cite{yu2022perpendicular, de2020flow, li2012formation, drossel2000phase, das2016phase, goswami2008nanoclusters, van2010hotspots, protein, veksler2007phase, gov2006dynamics, veksler2009calcium, kabaso2011theoretical, peleg2011propagating, fovsnarivc2019theoretical, yu2011early,  suarez2023reconstitution, litschel2024membrane, cagnetta2018active, chatterjee2007dynamics} and metastable states \cite{Chaudhuri2006, kwak2004driven, yu2022perpendicular} to disordered phases with unconventional dynamical universality classes \cite{fibonacci, nonlinear, popkovuniversality, chakrabortyUniversality, popkov2014superdiffusive}. Examples include clustering of GPI-anchored proteins in cell membranes due to coupling with the actomyosin network \cite{suarez2023reconstitution, litschel2024membrane, cagnetta2018active}, oscillatory transport in carbon nanotubes \cite{lee2010coherence}, and disordered systems displaying anomalous transport and mode-coupling phenomena \cite{popkov2004hydrodynamic, popkovexact, saito2021microscopic, miron2019derivation, chen2018exact, spohn2014nonlinear, mendl2013dynamic, spohn2015nonlinear, prakash2025exact}. The dynamically coupled behavior of cell membranes and the proteins associated with them \cite{peleg2011propagating, veksler2007phase, gov2006dynamics, veksler2009calcium, kabaso2011theoretical, fovsnarivc2019theoretical, legg2007n, bj2004bar} has attracted significant attention in recent years. These proteins both respond to and modify the local membrane curvature, leading to shape fluctuations. 

A fundamental driver in many such coupled nonequilibrium systems is the nature of the feedback between the interacting components. In general, this mutual coupling can be categorized into two types: aligned bias and reverse bias. An aligned bias occurs when a component acts on its environment in a manner that reinforces its own intrinsic dynamical preferences.  Conversely, a reverse bias arises when a component modifies its environment in opposition to its preferred direction of motion. In the presence of both aligned and reverse biases, the large scale  behavior of the system is often controlled by the relative strength of the two kinds of biases. It is generally expected that the stronger bias prevails. 






In this work we study a coupled driven system where two different species of hard core particles, `light' and `heavy' move on a fluctuating energy landscape. The heavy particles prefer to slide downward along local height gradient of the landscape, while the light particles prefer to slide upward. In addition, these particles modify the local height of the landscape by pushing it downward or pulling it upward. If the heavy particles, which themselves prefer to move downward, also push the landscape downward, we have the case of an aligned bias. But if they pull the landscape upward, it is reverse bias. Similarly, light particles pulling (pushing) the landscape upward (downward) correspond to aligned (reverse) bias. There are also vacancies in the system, representing parts of the landscape not occupied by any kind of particle species. This vacant parts of the landscape do not show any upward or downward bias and just undergo symmetric height fluctuations. This model is called light-heavy-vacancy or LHV model and was introduced in \cite{khamrai2026novel}. By changing the strengths of the aligned and reverse bias, a rich phase diagram was obtained which consists of various kinds of ordered and disordered phases \cite{khamrai2026novel}.

In this system aligned bias promotes long range order and reverse bias destroys it. A heavy particle which also pushes the landscape down, attracts other heavy particles due to their preference for lower height. This acts like a positive feedback and creates a large cluster of heavy particles and a deep valley of the landscape which holds the cluster. Aligned bias therefore creates ordering. On the other hand, if a heavy particle pulls the landscape upward, then even a small cluster of heavy particles would result in an increase in local landscape height, which in turn causes the heavy particles to go away from that neighborhood and the cluster gets dispersed. Reverse bias thus destabilizes local ordering. From this reasoning one expects strongest possible order when both particle species show aligned bias and completely disordered state when both of them show reverse bias. In the case when one species show aligned bias and the other one show reverse biased, general expectation is the final state would be determined by which bias is stronger. If aligned bias is stronger, long range order should prevail and if reverse bias is stronger, a disordered state is expected. Indeed that is the case when no vacancies are present in the system (LH model) \cite{chakrabortylarge,static,dynamic}.

However, for LHV model, which has vacancies, the above simple criterion does not hold \cite{khamrai2026novel}. This seems surprising since vacancies do not exert any kind of bias on the landscape and therefore, one would expect the competition between aligned and reverse bias should remain unaffected even when vacancies are present. But we have shown in an earlier study \cite{khamrai2026novel} that presence of vacancies significantly changes the criterion of order formation. In particular, LHV model allows long range order even when aligned bias is weaker than reverse bias. Only when both particle species show reverse bias, a disordered phase is obtained. As long as at least one species shows aligned bias, does not matter how weak that bias is, the system supports long range order \cite{khamrai2026novel}. In the next section we summarize these earlier results and explain the mechanism of order formation in this system.

In the present work, we investigate coarsening properties of these unusual kind of ordered phases. We specifically focus on those ordered phases where aligned bias is weaker than reverse bias. Just like existence of long ranger order is counter-intuitive here, we find the coarsening process itself is highly unconventional for these phases. Our numerical simulations show that the particle coarsening is algebraic in nature but the coarsening length scale grows with two different exponents at early times and late times. The local ordered domains which develop at early times in certain special zones of the configuration, eventually become unstable and they disintegrate. Instead, new domains emerge elsewhere in the system which coalesce to coarsen further. In a conventional coarsening process small ordered domains form which merge and the size increases. Such merger keeps happening and the length scale grows with a single power law exponent at all times. On the contrary, in our system the early time domains become unsustainable and break up after some times. New domains form at later times whose merger follows a different growth law. The landscape coarsening shows even more surprising trends. It shows distinct non-monotonic behavior with time. We show that this happens due to presence of three normal modes which move through the system like a traveling wave. Using linear hydrodynamics we show that the propagation velocities of these modes are given by eigenvalues of the current Jacobian matrix which we calculate within mean field approximation.

The remainder of the paper is organized as follows. In Sec. \ref{sec:model} we introduce the LHV model and give a brief overview of the phase diagram. In Sec. \ref{sec:ParCoar} we present our results for particle coarsening and in Sec. \ref{sec:SurOsci} we discuss surface coarsening. Sec. \ref{sec:Conclu} contains few concluding remarks. Details of hydrodynamic calculations are presented in Appendix \ref{App:KinWave}.

\section{LHV Model and Phase Diagram } \label{sec:model}

\begin{figure}[H]
\centering
\includegraphics[scale=0.37]{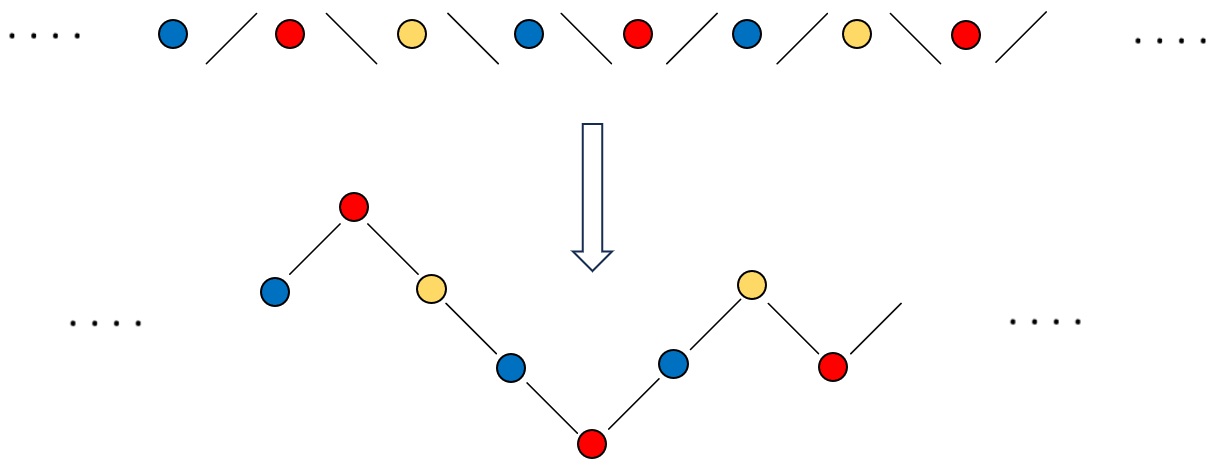}
\caption{Mapping a given configuration onto a height model. Red, yellow, and blue circles represent heavy particles, light particles, and holes; $\slash$ and $\backslash$ denote upslope and downslope bonds respectively.} 
\label{fig:hconf}
\end{figure}

Our model is defined on a one-dimensional periodic lattice in which each site may be occupied either by a heavy $(H)$ particle, a light $(L)$ particle, or may remain vacant $(V)$. A vacant site is also referred to as a hole. The particles obey hardcore exclusion, implying that no lattice site can be simultaneously occupied by more than one particle. The bonds connecting neighboring lattice sites $j$ and $(j+1)$ can adopt two possible orientations, $\pm \pi/4$, represented by the variable $\tau_{j+1/2}$. A bond with orientation $\pi/4$ corresponds to an upslope bond, while one with orientation $-\pi/4$ corresponds to a downslope bond; these are denoted by $\tau_{j+1/2}=1$ and $\tau_{j+1/2}=-1$, respectively. We further introduce a height field associated with each lattice site, defined as $h_i = \sum_{j=1}^{i-1} \tau_{j+1/2}$. In schematic representations of lattice configurations, an upslope bond is depicted by the symbol $\slash$, whereas a downslope bond is represented by $\backslash$. Site occupancies are illustrated using colored circles: red for an $H$ particle, yellow for an $L$ particle, and blue for an empty site. Figure \ref{fig:hconf} presents a representative configuration containing different particle species on a fluctuating surface with a spatially varying height profile.  

\begin{figure}[H]
\centering
\includegraphics[scale=0.25]{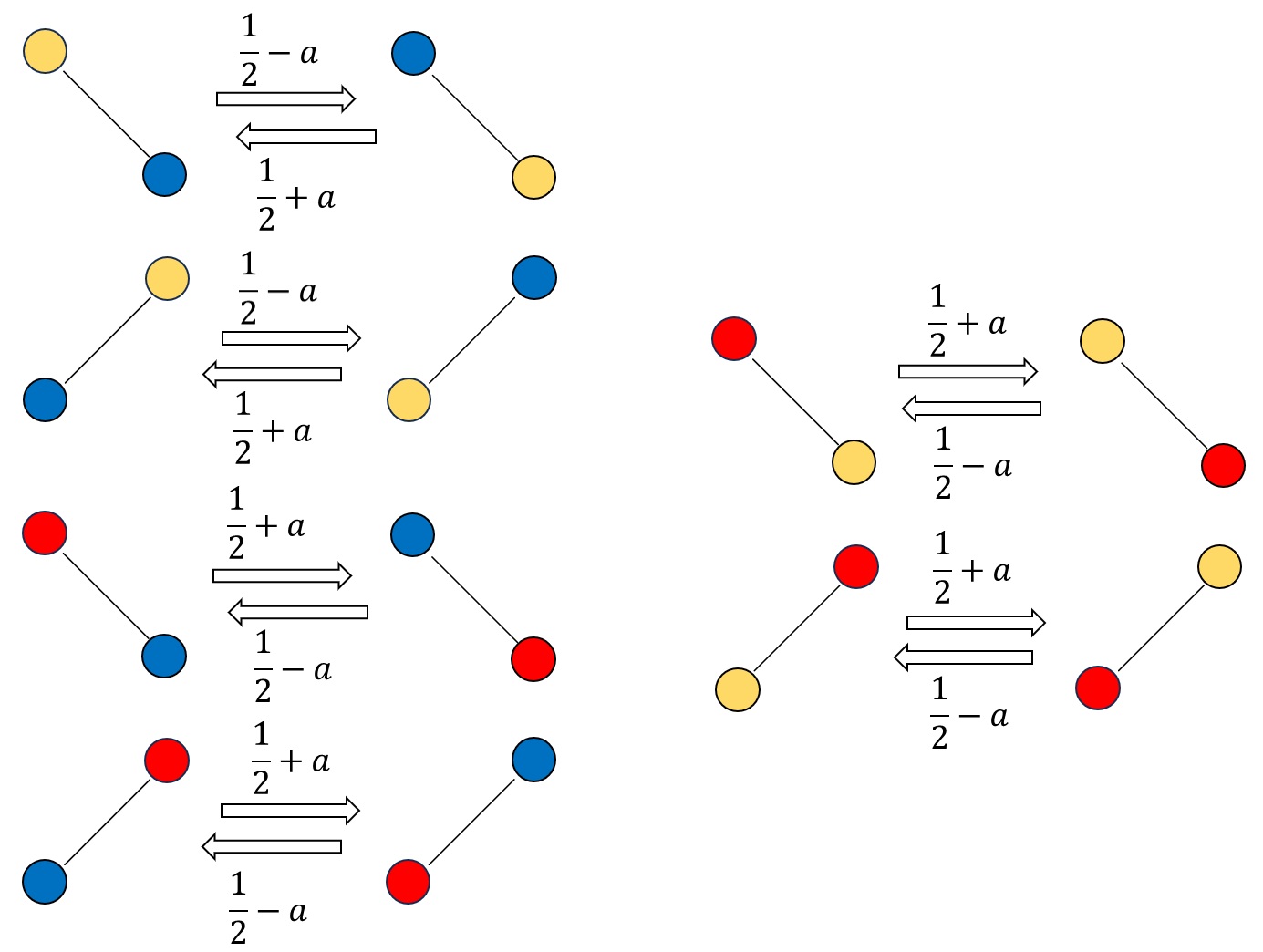}
\caption{Particle transition rates under the condition $0 < a \leq \frac{1}{2}$.} \label{fig:ParUpdate}
\end{figure}

Particle exchange occurs between neighboring lattice sites by sliding along the intervening bonds, with their transition rates coupled to the local surface slope. Specifically, $H$ particles preferentially slide down the surface, while $L$ particles favor upward hopping. These preferences govern the stochastic rates at which an $H$ ($L$) particle undergoes a local position exchange with a neighboring $L$ ($H$) particle or an empty site. As an example, any exchange that displaces an $H$ particle to a site at a lower height, replacing an $L$ particle or a hole, is accepted with a probability of $(\frac{1}{2}+a)$, whereas the reverse process occurs with a reduced probability of $(\frac{1}{2}-a)$. Here, the driving parameter satisfies $0 < a \leq \frac{1}{2}$. The allowed particle configurations and updates are illustrated in Fig. \ref{fig:ParUpdate}. We have taken $a = 0.4$ throughout this study.


\begin{figure}[H]
\centering
\includegraphics[scale=0.28]{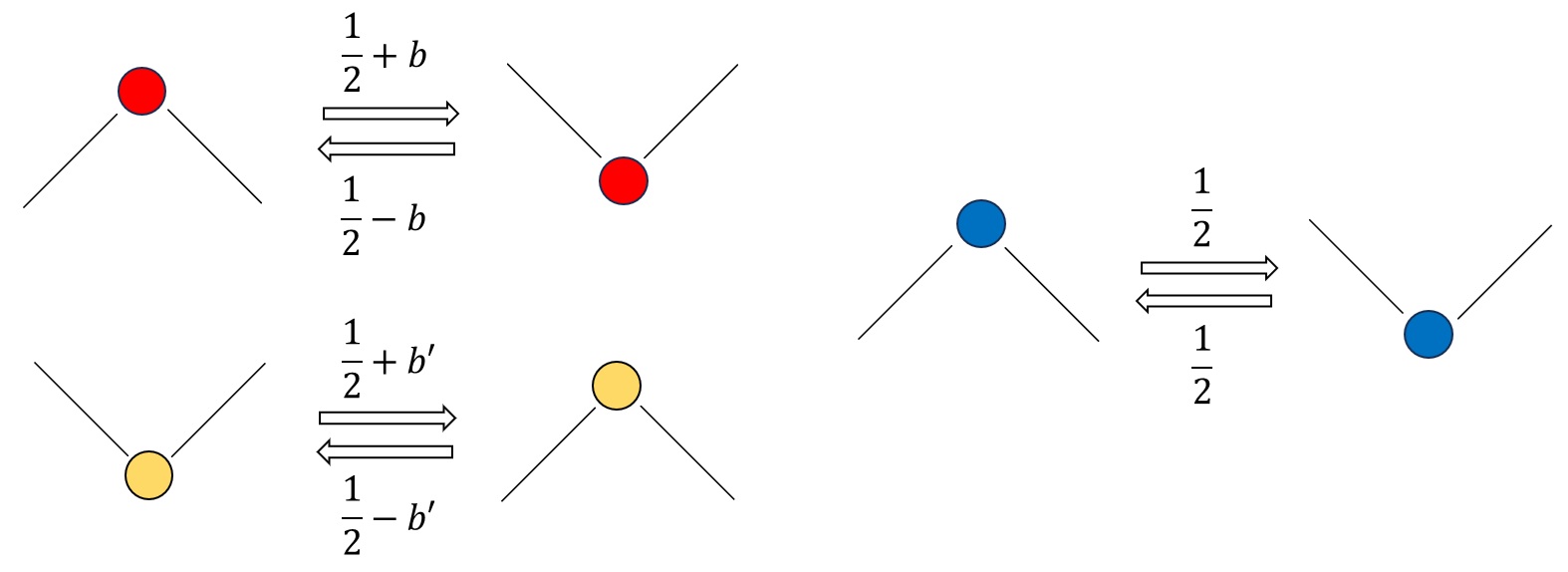}
\caption{Local landscape transition rules for $H$ particles, $L$ particles, and holes occupying local hills and valleys.} \label{fig:smoves}
\end{figure}

The model incorporates a bidirectional coupling where surface dynamics and particle movement mutually influence one another; specifically, the local occupancy of particles dictates surface evolution, which in turn alters particle kinetics. Surface configurations modify through the orientation exchange of adjacent bonds. In this framework, a local ``hill'' is defined by an upslope bond preceding a downslope bond, whereas the inverse sequence constitutes a local ``valley.'' The transition rates between a hill and a valley are governed by the occupancy of the intermediate site. As illustrated in Fig. \ref{fig:smoves}, a hill containing an $H$ particle transforms into a valley with probability $(\frac{1}{2}+b)$, while the inverse process occurs with probability $(\frac{1}{2}-b)$. Consequently, when $b > 0$, the hill-to-valley transition is favored than valley-to-hill transition, meaning $H$ particles effectively ``push down'' the surface. Therefore $b>0$ corresponds to aligned bias for $H$. Conversely, $b<0$ implies that $H$ particles elevate or ``pull up'' the surface, which corresponds to reverse bias. A similar condition applies to $L$ particles: a valley holding an $L$ particle flips to a hill with probability $(\frac{1}{2}+b')$ and flips back with probability $(\frac{1}{2}-b')$. Thus, a positive $b'$ causes $L$ particles to pull the surface upward, whereas a negative $b'$ results in a downward push. Since $L$ themselves prefer to move upward, positive (negative) $b'$ therefore represent aligned (reverse) bias. The forward and reverse transitions between hill and valley become symmetric when the intervening site is vacant, each occurring with a baseline probability of $\frac{1}{2}$ (see Fig. \ref{fig:smoves}). The parameters $b$ and $b'$ are both constrained within the interval $[-\frac{1}{2},\frac{1}{2}]$. 

In the absence of any vacant sites, a closely related model was introduced in \cite{chakrabortylarge} as LH (light-heavy) model and was extensively studied in \cite{dynamic, static, chakrabortyUniversality, lightheavy, khamrai2024effect, prakash2025exact, prakash2026critical}. The present LHV model is defined on a ring lattice with $N$ sites and $N$ bonds out of which $N/2$ bonds are upslope and the remaining $N/2$ are downslope. Let $N_H$ and $N_L$ represent the total populations of $H$ and $L$ particles, respectively; the total number of vacancies is consequently given by $N - N_H - N_L$. We define the respective particle densities as $\rho_{H} = N_{H}/N$ and $\rho_{L} = N_{L}/N$. Unless specified otherwise, the numerical data presented here are obtained using equal densities $\rho_H = \rho_L = 1/3$. However, all our conclusions hold for any finite concentrations of $H$ particles, $L$ particles, and vacancies. All our data have been averaged over a minimum of $10^5$ independent histories.

\begin{figure}[H]
\centering
\includegraphics[scale=0.85]{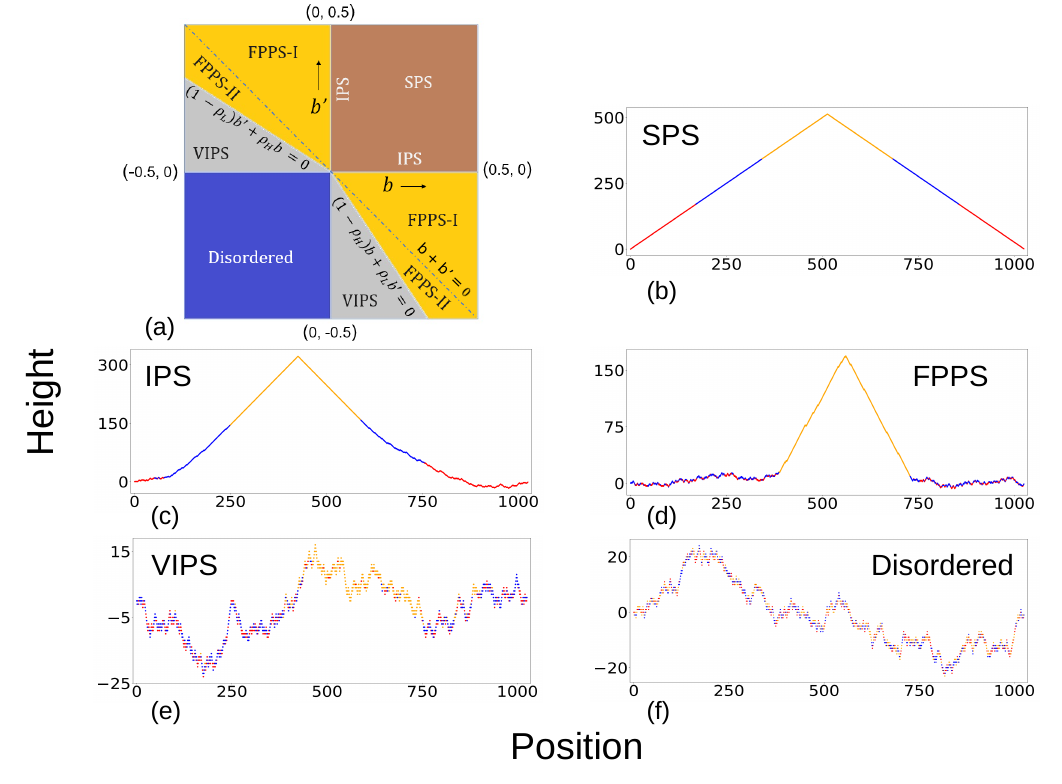}
\caption{Phase diagram in steady state of the LHV model and typical configurations corresponding to the different phases. The three colors---red, yellow, and blue---denote $H$ particles, $L$ particles, and vacancies, respectively. The parameter values used for the representative configurations are: (b) SPS, $b=b'=0.4$; (c) IPS, $b'=0.4$ and $b=0$; (d) FPPS, $b'=0.4$ and $b=-0.1$; (e) VIPS, $b'=0.18$ and $b=-0.4$; and (f) disordered phase, $b=b'=-0.4$. All configurations correspond to a system of size $N=1026$, with equal densities $\rho_H=\rho_L=1/3$, and $a=0.4$. The phase diagram remains valid for arbitrary nonzero densities of both particle species and vacancies in the thermodynamic limit $N \rightarrow \infty$.} \label{fig:OccProb}
\end{figure}

In an earlier work \cite{khamrai2026novel} we had derived the phase diagram of LHV model. We briefly describe the phases here. Since aligned bias promotes ordering and reverse bias destroys it, highest possible ordered state is expected when both $b$ and $b'$ are positive. In Fig. \ref{fig:OccProb}b a representative configuration is shown where all upslope bonds are completely phase separated from downslope bonds, giving rise to a deep valley and a large hill in the surface. All $H$ form a single cluster and occupy the valley, all $L$ can be found at the hill and all vacancies are present at the interface between $L$ and $H$ clusters. This is called Strong Phase Separation (SPS).

When one of the coupling parameters $b,b'$ is positive and the other one is zero, one particle species show aligned bias and the other species show no bias. In this case all particles and vacancies still phase separate like SPS but the surface does not show perfect order. Only one part of the surface that is beneath the cluster of particles species showing aligned bias, shows phase separation between upslope and downslope bonds to form a sharp hill (valley) for $b'>0$ ($b>0$). But the remaining part of the surface flattens out in the shape of a parabola where the probability to find an upslope bond varies linearly with distance and the gradient scales as $1/N$. This gradient also generates a surface current with similar scaling. This phase is known as Infinitesimal current with phase separation (IPS). In Fig. \ref{fig:OccProb}c we show a representative configuration for the case $b'>0$ and $b=0$.

When $b$ and $b'$ have opposite signs, one particle species shows aligned bias and the other one shows reverse bias. For $b+b'>0$ the aligned bias remains stronger than reverse bias. In this case the aligned species phase separate to form a single cluster and the surface underneath forms a large hill (or valley) which is not perfectly sharp but has a rugged shape (Fig. \ref{fig:OccProb}d) due to minority up-slope (down-slope) bonds interspersed within majority down-slope (up-slope) region. The remaining part of the surface is completely disordered and contains a random mixture of vacancies along with the particle species with reverse bias. A finite surface current flows through the system. This phase is named finite current with partial phase separation (FPPS-I) and we distinguish it from FPPS-II for which $b+b'<0$. Although typical steady state configuration for FPPS-II phase looks similar to FPPS-I, there is an important distinction between the two phases. The aligned bias is weaker than reverse bias for FPPS-II but long range order still prevails in the system. The species showing stronger reverse bias remains mixed with the vacancies, which have zero bias. The effective bias exerted by this mixture balances the weaker aligned bias of the phase separated species.

However, when reverse bias becomes much stronger than the aligned bias, then even after mixing with the vacancies the effective reverse bias wins over the aligned bias. Within mean field theory it can be shown that the condition for this is $(1 - \rho_Lb') + \rho_Hb < 0$ with $b' > 0$ and $(1 - \rho_Hb) + \rho_Lb' < 0$ for $b>0$ \cite{khamrai2026novel}. In this case, the system still does not become completely disordered. The particle species showing aligned bias still phase separates but instead of forming a single cluster, a small fraction of the other particles species is mixed with it. The underlying surface takes the shape of a plateau, instead of a large hill or valley. This phase is called vacancy induced phase separation (VIPS). 

For both $b,b' < 0$ no long range order is found in the system. Therefore, as long as at least one species shows aligned bias, that species always phase separates, even when reverse bias has a significantly larger magnitude. This unusual kind of ordering can be seen only in FPPS-II and VIPS phase. Below we study the kinetics of this phase ordering.

\section{Two different Growth Regimes in Particle Coarsening} \label{sec:ParCoar}

\begin{figure}[H]
\centering
\includegraphics[scale=1]{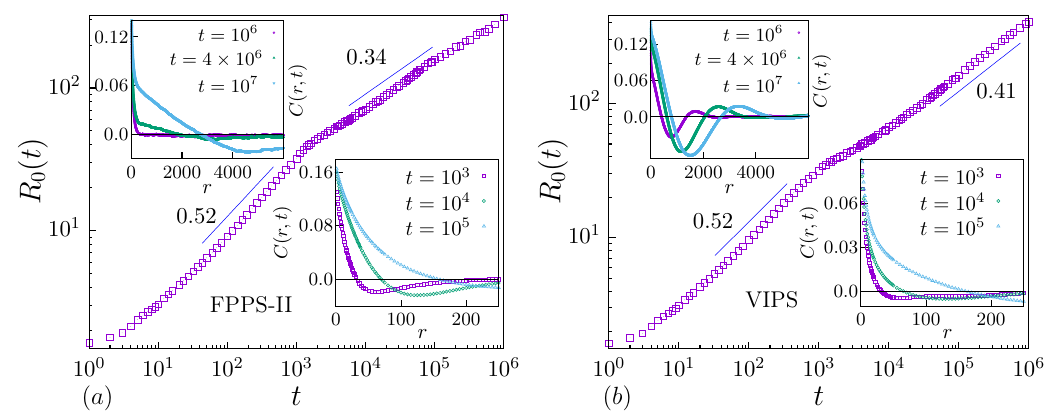}
\caption{Variation of the coarsening length scale $R_0(t)$ with time for (a) the FPPS-II phase ($b' = 0.3$) and (b) the VIPS phase ($b' = 0.18$), for system size $N = 16416$ and $b = -0.4$. In both phases $R_0(t)$ grows as a power law, with exponent $\simeq 0.5$ at early times ($t \lesssim 10^3$), crossing over to a smaller exponent at late times; $ \simeq 0.34$ for FPPS-II (left) and $ \simeq 0.41$ for VIPS (right). Lower insets: the two-point density-density correlation function, showing the outward shift of its zero crossing as $t$ increases. Upper insets: $C(r,t)$ at late times, showing the qualitatively different late-time shape of the correlation function in the two phases.} \label{fig:TwoPoi}
\end{figure}

In this section we focus on the coarsening behavior of the particle species in FPPS-II phase and VIPS phase. From our discussions in the previous section it follows that in both these phases the aligned bias is weaker than the reverse bias and a long range order in general is not expected, but for the presence of vacancies. To see how the long range order develops for the particles, we start with a randomly disordered configuration and monitor how growing length scales in particle correlations emerge with time. We measure $C(r,t) = \langle \eta_L(j,t) \eta_L(j+r,t) \rangle - \rho_L^2  $, where $\eta_L(i,t)$ takes the value $1$ if site $i$ at time $t$ is occupied by an $L$ particle, otherwise $\eta_L(i,t)=0$. In Fig. \ref{fig:TwoPoi}a and b insets we plot $C(r,t)$ for different values of $t$. We find that at late times $C(r,t)$ shows different qualitative shape for FPPS-II and VIPS phases. For FPPS-II $C(r,t)$ shows sharp fall at small $r$ and then decreases more slowly with $r$ to cross zero once and then becomes negative at large $r$. But for VIPS it crosses zero multiple times and finally levels off at zero for large $r$, as shown in top left inset of Fig. \ref{fig:TwoPoi}b. All simulation data presented in this paper have been averaged over at least $10^4$ histories.

To extract the coarsening length scale from $C(r,t)$, we define $R_0(t)$ as the distance at which first zero-crossing of $C(r,t)$ occurs at a given $t$. In the main plots of the same figures we show the variation of $R_0(t)$ with time. Our data show that $R_0(t)$ grows with $t$ as a power-law, but the coarsening exponent takes two distinctly different values for large and small $t$. Both for FPPS-II and VIPS phases (Fig. \ref{fig:TwoPoi}a and b, respectively) the coarsening exponent at small times ($t \lesssim 10^3$) is close to $0.5$ but at long times, the exponent changes to a smaller value. Our numerical data indicate particle coarsening in FPPS-II phase progresses with an exponent $\simeq 0.34$ at large times, while for VIPS phase the exponent is $\simeq 0.41$. Having two distinctly different growth regimes during coarsening is rather unusual. We explain this effect below.


Starting from a disordered configuration, the $L$ particles locally phase separate by displacing the $H$ and the vacancies. Consider a small zone where $L$ and $H$ are mixed together. Both these species push the surface up and $H$ particles do so at a higher rate. So the height of the surface around $H$ tends to go up but $H$ themselves prefer to roll down to low-lying parts of the surface. This prevents any local phase separation between $L$ and $H$ within this zone. Next consider another small zone where $L$ are mixed with vacancies. Here local phase separation happens more easily: $L$ push the surface up and create small hills around themselves and localize there by displacing the vacancies which play an inert role. This means initial phase separation of $L$ particles happens within small zones or pockets  where $L$ are mixed with vacancies. The coarsening behavior in these pockets is akin to that of IPS phase of $LH$ model which is known to show diffusive coarsening \cite{chakrabortylarge, dynamic}. This explains the early time exponent observed in Fig. \ref{fig:TwoPoi}. Note that the surface current in these zones with $L$ and vacancies are being driven by $L$ since the vacancies do not contribute to the surface current. This current is much smaller than other segments of the system where $H$ and $L$ both are present. Due to this mismatch in surface currents the phase separation process between $L$ and $V$ can not sustain beyond a certain time. Eventually $H$ start entering these  locally phase ordered pockets of $L$ and $V$ and displacing the $L$ from there. These $L$ particles then join other $L$ particles which were earlier mixed with $H$ in the initial configuration. $R_0(t)$ in this time regime shows a slower growth. The crossover time $t \sim 1000$ between the early time diffusive growth and the late time slower growth does not depend on $N$ (data not shown here).

\begin{figure}[H]
\centering
\includegraphics[scale=0.85]{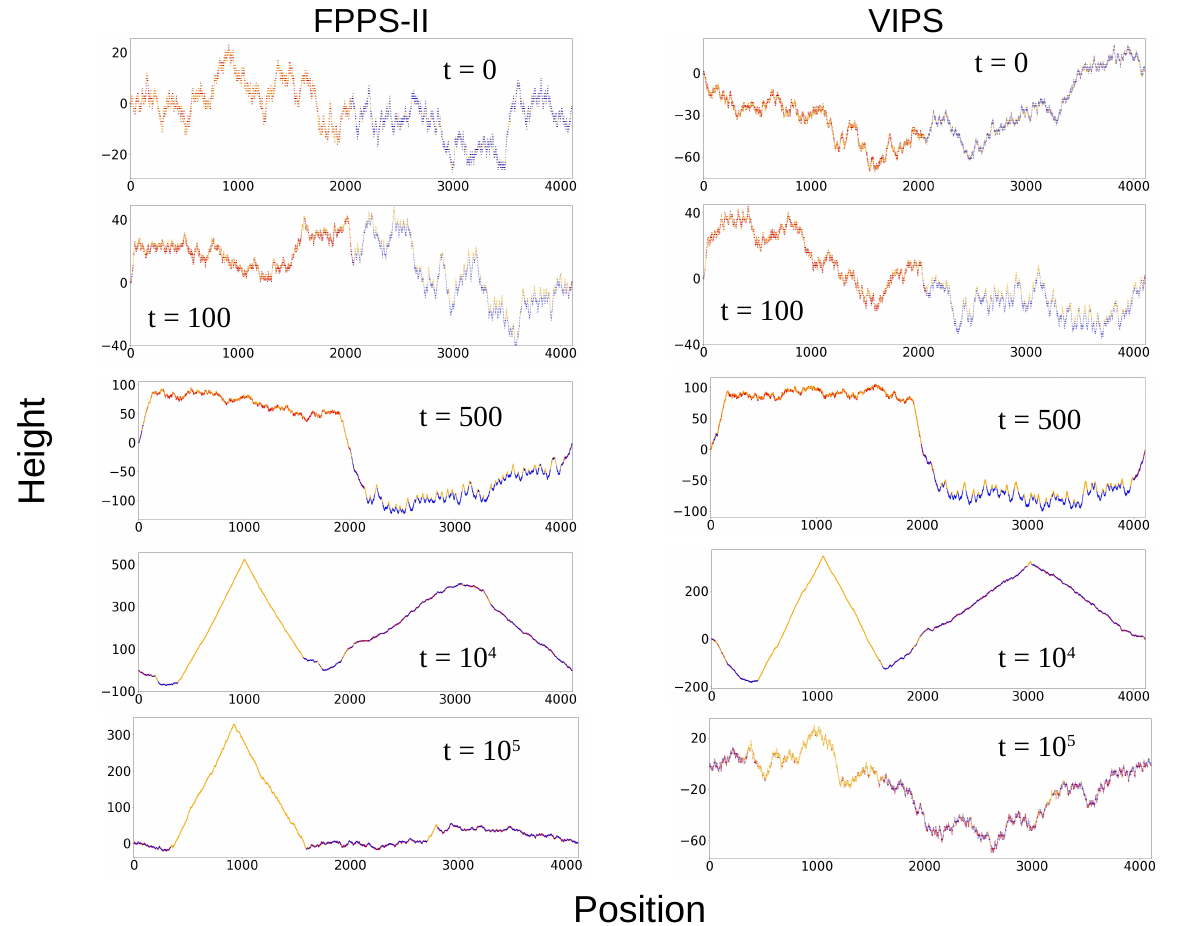}
\caption{Snapshots capturing a single realization of the coarsening mechanism under specific initial conditions. Simulation parameters are fixed at $b' = 0.3$ for FPPS-II phase (left) and $b' = 0.18$ for VIPS phase (right), with $b = -0.4$ and $N = 4104$ for both the cases. Small clusters of light particles are visible in the right half of the configuration at $t = 100$. At $t = 500$, the mismatch of surface current between left and right sector is reflected as the height difference between these two sectors. Around $t = 10^4$, we find two macroscopic hills for both FPPS-II and VIPS phase. This effect is arising due to choice of special initial condition. When we start from a random initial condition, a macroscopic hill appears only in the late time coarsening regime of the FPPS-II phase.} \label{fig:CoarSpeCon}
\end{figure}

The above coarsening mechanism can be clearly visualized if we start with a special initial configuration where the surface is completely disordered, and one half of the surface holds a random mixture of $L$ and $H$, while the other half contains $L$ and vacancies in a mixed state. In Fig. \ref{fig:CoarSpeCon} we show the time evolution for FPPS-II and VIPS phase. Indeed we find initial phase separation happens between $L$ and vacancies, while the segment with $L$ and $H$ remains disordered. The mismatch in local surface current can also be clearly seen from the height difference created between these two segments. Since both $L$ and $H$ are pushing the surface up, the surface beneath the $L$-$H$ mixture moves up at a fast rate. On the other hand, in the partially phase separated region only $L$ apply upward bias and that too at a rate lower than the $H$. As a result the surface in this sector moves up at a slower rate. The height difference created as a result of this eventually becomes unsustainable. The $H$ then start rolling down from the higher region and $L$ from the lower region start climbing up. This gives rise to the late stage of the coarsening process 

\begin{figure}[H]
\centering
\includegraphics[scale=1]{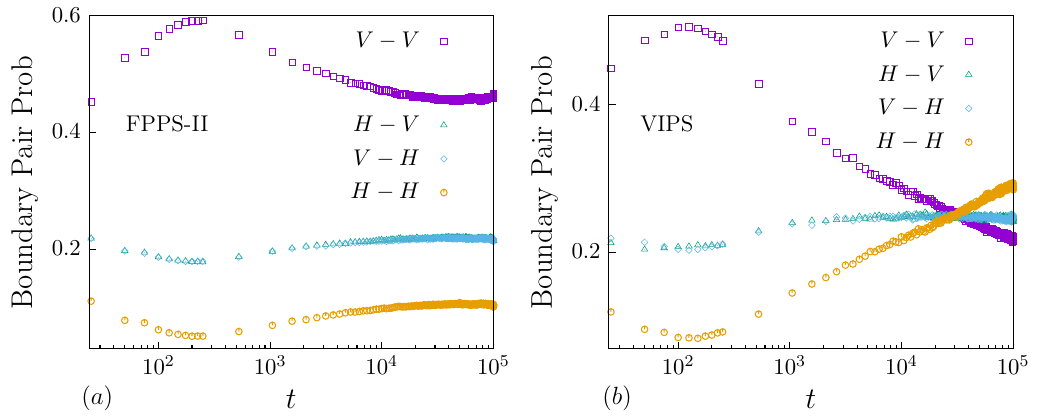}
\caption{Variation of probabilities of the four possible particle configurations at the two ends of the largest light-particle cluster during coarsening, starting from a random initial condition. Left: FPPS-II phase with $b'=0.3$; right: VIPS phase with $b'=0.18$. In both cases, $b=-0.4$ and $N=16416$.
} \label{fig:ParLenScale}
\end{figure}

The time-evolution shown in Fig. \ref{fig:CoarSpeCon} clearly illustrates the mechanism of coarsening process for a specially chosen initial configuration. The same mechanism is at work even for a randomly disordered initial configuration, although just by looking at a typical time-evolution it may not become clear in this case because of fluctuations present. To independently verify that the same mechanism is at play even for a random disordered initial configuration, we measure the following quantity. At any time $t$ consider the largest $L$ cluster present in the system. At the two boundaries of this cluster one can have either an $H$ or a vacancy $V$. We measure how the probabilities associated with the four possible boundary-pair configurations change with time. Our data in Fig. \ref{fig:ParLenScale} show that, among the four possible configurations, largest $L$ particle cluster having vacant sites at both ends is the most probable at early times of the coarsening regime, both in the FPPS-II and VIPS phase. The configuration with $H$ particles at both ends of the largest $L$ particle cluster is the least probable. This confirms the early time local phase separation between $L$ and vacancies. For VIPS phase, this trend gets reversed at some intermediate time as in the steady state, there is a small fraction of $H$ particles residing within the $L$-particle region \cite{khamrai2026novel}.

\section{Oscillations in Surface Coarsening} \label{sec:SurOsci}

\begin{figure}[H]
\centering
\includegraphics[scale=1]{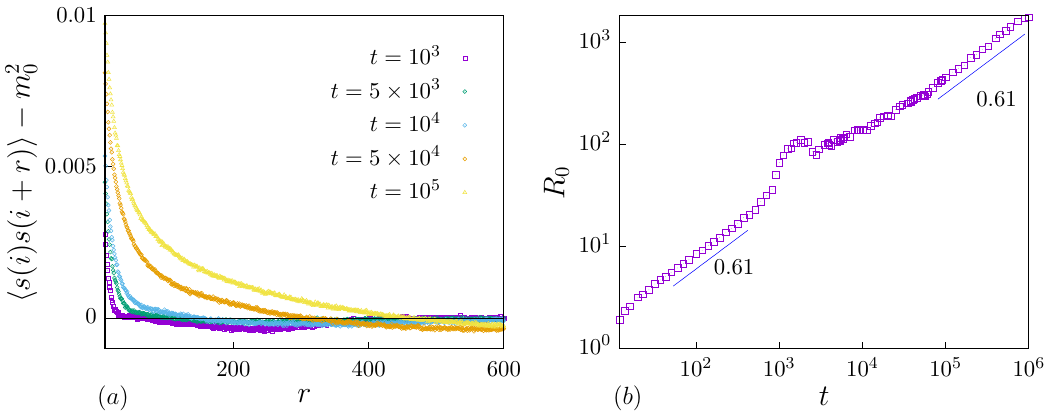}
\caption{Two-point density correlation function of upslope bonds plotted against separation $r$ at several times (left) and non-monotonic behavior of the surface coarsening length scale (right) for FPPS-II phase. Simulation parameters are fixed at $b = -0.4$, $b' = 0.25$, and $N = 16416$.} \label{fig:SlSlR0}
\end{figure}

In the FPPS-II phase the $L$ cluster occupies a large macroscopic hill in steady state. This large hill is a result of partial phase separation between upslope and downslope bonds. To study how this phase separation progresses with time, we monitor the two-point correlation function between bond orientation: ${C_s} (r,t)= \langle s(i,t) s(i+r,t) \rangle - \langle s(i,t) \rangle^2 $, where $s_i(t)$ takes the value $1$ (or $0$) if the bond between site $i$ and $(i+1)$ is an upslope (downslope) bond at time $t$. Using the notation introduced in Sec. \ref{sec:model} $s_i(t)=(1+\tau_{i+1/2})/2 $. Since the dynamics conserves total number of upslope and downslope bonds, $\frac{1}{N}\sum_i s(i,t)=m_0$ is conserved. Here we have considered $m_0 = 1/2$. In Fig. \ref{fig:SlSlR0}a we plot ${C_s} (r,t)$ as a function of $r$ for few $t$ values. The distance at which ${C_s} (r,t)$ crosses zero for the first time is denoted as $R_0(t)$ which we identify as the coarsening length scale. In Fig. \ref{fig:SlSlR0}b we show the time variation of $R_0(t)$ and find a power law growth for most values of $t$. The most surprising feature of Fig. \ref{fig:SlSlR0}b is the non-monotonic behavior of $R_0(t)$ at intermediate times. The coarsening length scale in a finite system is expected to increase with $t$ and saturate as steady state is reached. It is highly unusual for a coarsening length scale to decrease while the system is relaxing towards phase separated state. To the best of our knowledge, such behavior has never been reported in any phase ordering system before.

\begin{figure}[H]
\centering
\includegraphics[scale=1]{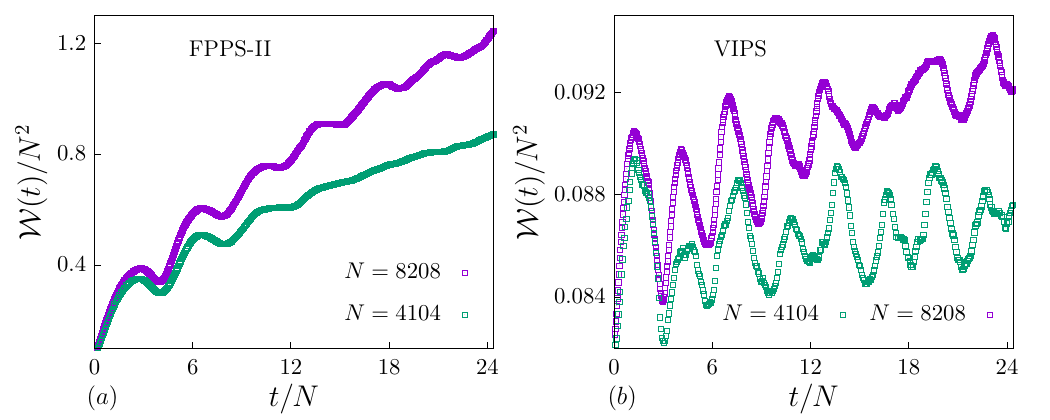}
\caption{Variation of interface width in the coarsening regime of the FPPS-II (left) and VIPS phase (right) for two different system sizes. The parameters are $b' = 0.3$ for FPPS-II phase and $b' = 0.18$ for VIPS phase, with $b = -0.4$ for both of these cases. Periodic oscillations are observed where time period scales with system size.} \label{fig:CoarSurWidth}
\end{figure}

In the VIPS phase no macroscopic hill is present in steady state. Instead, an elevated plateau is formed which holds the $L$ particles. We define the surface width as $\mathcal{W}(t) = \langle \sum_{i=1}^{N} (h_i(t) - {\bar h(t)})^2\rangle/N$, where $h_i(t)$ is the height of the surface at site $i$ at time $t$ and ${\bar h(t)}$ is the average height across all sites. Compared to a disordered surface, the width is significantly higher for a surface with a plateau. Therefore, to monitor relaxation of the surface in VIPS phase we measure $\mathcal{W}(t)$ as a function of time starting from a disordered random configuration. Note that in FPPS-II phase the surface width is even larger because of the large hill present in steady state. In Fig. \ref{fig:CoarSurWidth} we show temporal variation of $\mathcal{W}(t)$ for both these phases. Surprisingly, $\mathcal{W}(t)$ shows periodic oscillations along with an increasing trend. Our data also show that the time-period of oscillation scales with system size. While for VIPS phase oscillation patterns are a bit more haphazard than FPPS-II phase, the presence of maxima and minima of $\mathcal{W}(t)$ at periodic intervals of time can clearly be seen. 

Thus our data in Figs.  \ref{fig:SlSlR0} and \ref{fig:CoarSurWidth} indicate that coarsening process of surface progresses in a non-monotonic manner: while showing an overall increase, both $R_0(t)$ and ${\mathcal W}(t)$ also sometimes decrease with $t$. This unusual behavior can be explained from the presence of kinematic waves \cite{das2001weak, chatterjee2007dynamics, gupta2007tagged, gupta2007finite} in the system. In Appendix \ref{App:KinWave} we present detailed calculations. These calculations show that any fluctuations in the local density of upslope bonds can be decomposed into three normal modes which satisfy traveling wave equations with propagation velocities given by
\begin{equation} 
\lambda = 0, \pm \sqrt{ (-2a) \left [ \rho_H(1-\rho_H)b + \rho_L(1- \rho_L)b' \right ] } 
\end{equation}
{\sl i.e.} one mode remains stationary and two other modes move through the system with equal and opposite velocities. This effect has been clearly demonstrated in  Fig. \ref{fig:momt} where space-time correlation of upslope bonds is plotted. In our measurement of height fluctuations the presence of these traveling waves are felt too. The fluctuation generated at a point gets carried away by these  traveling waves and surface width ${\mathcal W}(t)$ keeps increasing ballistically ($\sim t^2$). Since we have periodic boundary condition, two oppositely moving waves with speed $\pm \lambda$ meet after a time-interval $T=N/2 \lambda$ when ${\mathcal W}(t)$ reaches its first minimum. We compare the position of the first minimum in Fig. \ref{fig:CoarSurWidth} with mean field prediction. For FPPS-II phase $T/N$ observed from simulations is $\simeq 4.08$ while mean field predicts $3.85$. For VIPS phase simulation gives $T/N \simeq 2.96$ and mean field yields $2.53$. Thus the quantitative prediction of mean field theory does not work very well. This is not unexpected since mean field ignores the correlations that build up in the system during coarsening phase. However, even with lack of quantitative match, our theory successfully explains the origin of oscillations observed during coarsening process.

\section{Conclusions} \label{sec:Conclu}

In this work we have explored unusual phase ordering dynamics in a coupled driven system consisting of different particle species along with vacancies moving on fluctuating energy landscape. We have specifically focused on those cases for which order-inducing interaction is weaker than order-destroying interaction. Although long range order is generally not expected in such cases, we find presence of vacancies still takes the system towards ordered phases. However, the dynamics of ordering here is very much different from conventional phase ordering systems. Instead of small ordered domains coalescing and creating larger domains, in our model the ordered domains which are formed at short times become unstable at large times and break apart. New ordered domains emerge at late times which finally guide the system towards an ordered state. As a result, the particle coarsening progresses with two different power law exponents at early and late times. The landscape coarsening shows non-monotonic variation of characteristic length scale which is a direct outcome of breaking down of earlier ordered domains at intermediate times. We also show presence of traveling wave structures in the coarsening phase which gives rise to interesting time-periodic oscillations in the height fluctuations of the landscape.

Throughout this study, we have focused on only those phases where formation of long range order itself is counter-intuitive. In other phases of LHV model where order-inducing aligned bias is actually stronger than order-destroying reverse bias, the coarsening dynamics remains similar to those found in conventional ordered system. We have verified (data not shown here) that in these phases both particle coarsening and landscape coarsening progress in an usual manner and no non-monotonicity or sudden change in coarsening exponent with time are observed. Therefore it is essentially the interplay between weaker aligned bias and stronger reverse bias and how presence of vacancies tilt the competition in favor of weaker aligned bias, which is responsible for the surprising phase ordering kinetics reported here. It will be of interest to see how general these results are. Whether it is possible to see similar unusual coarsening dynamics in other kinds of systems with competing interactions, remains an open question.

It may be possible to design experiments to test some of our conclusions. Membrane proteins which induce curvatures in the cell membrane and also follow those curvatures may be used for this purpose \cite{saarikangas2009molecular, mim2012membrane}. Sometimes there can even be mismatch between intrinsic protein curvature and induced membrane curvature which may be used as an example of reverse bias \cite{millard2005structural, mattila2007missing}. By considering such a system of membrane and different kinds of proteins one can track the dynamics of formation of ordered structure.

\section{Acknowledgements}
We acknowledge useful discussions with Mustansir Barma. CK acknowledges research fellowship (Grant No. 09/0575(12571)/2021-EMR-I) from the Council of Scientific and Industrial Research (CSIR), India. SC acknowledges support from Anusandhan National Research Foundation (ANRF), India (Grant No: CRG/2023/000159).

\appendix

\section{Presence of Kinematic Waves in Coarsening Phase}\label{App:KinWave}

In our model there are three conserved quantities: number of $L$, $H$ particles and number of upslope bonds. Let us denote by $\rho_l(x,t)$, $\rho_h(x,t)$ and $m(x,t)$ the local densities of $L$, $H$ and upslope bonds, respectively. One can write down continuity equations for each of them. These equations will be coupled, {\sl i.e.} local current of one conserved species will in general depend on the local density of other conserved species. For example, current of upslope bonds $J_m$ can be formally written as 
\begin{equation}
J_m = P(/H\backslash)(E + b) - P(\backslash H/)(E - b) + P(/L\backslash)(E - b') - P(\backslash L/)(E + b') 
\end{equation}
which involves three point correlation functions like $P(/H\backslash)$ denoting the probability of a microscopic hill occupied by an $H$ particle. We do not have exact expression for these correlation functions and use mean field approximation to factorize them. Although mean field may not be a good approximation in a phase ordering system where long range correlations emerge with time, we show below that it can still provide insight into the non-monotonic variation of coarsening length scale and surface width. The mean field expression for $J_m$ becomes 
\begin{equation}
    J_m = 2m(1-m)(\rho_hb-\rho_l b').
\end{equation}
In a similar manner we can write for local currents for $L$ and $H$
\begin{eqnarray*}
J_l & = & -2a \rho_l (1-\rho_l) (1-2m) \\   J_h & = & 2a \rho_h (1-\rho_h) (1-2m)
\end{eqnarray*}
The continuity equations read 
\begin{eqnarray*}
{\partial_t} \rho_l(x,t) + {\partial_x} J_l(x,t) &= & 0 \\
{\partial_t} \rho_h(x,t) + {\partial_x} J_h(x,t) &= & 0 \\
{\partial_t} m(x,t) + {\partial_x} J_m(x,t) &= & 0 
\end{eqnarray*}
Here, the space-time dependence of currents comes via local densities and currents do not show any explicit dependence on $x$ and $t$. Assuming the local densities vary slowly enough in space, we perform hydrodynamic expansions  $\rho_l(x,t) = \rho_L + \tilde{\rho_l}(x,t)$, $\rho_h (x,t) = \rho_H + \tilde{\rho_h}(x,t)$ and $m(x,t) = m_0 + \tilde{m}(x,t)$ and retain terms up to linear order in $\tilde{\rho_l}$, $\tilde{\rho_h}$ and $\tilde{m}$.

The current Jacobian matrix $A$ has elements $A_{\alpha \beta} = \partial J_\alpha / \partial \rho_\beta$ and can be written as 
\begin{equation}
A=
\begin{bmatrix}
2a(1-2m_0)(1-2\rho_H)
&
0
&
-4a\rho_H(1-\rho_H)
\\[2mm]
0
&
-2a(1-2m_0)(1-2\rho_L)
&
4a\rho_L(1-\rho_L)
\\[2mm]
2m_0(1-m_0)b
&
-2m_0(1-m_0)b'
&
2(b\rho_H-b'\rho_L)(1-2m_0)
\end{bmatrix}.
\end{equation}
and for $m_0 = 1/2$ it takes the form
\begin{equation} 
A= \begin{bmatrix} 0 & 0 & -4a\rho_H(1-\rho_H) \\[2mm] 0 & 0 & 4a\rho_L(1-\rho_L) \\[2mm] \dfrac{b}{2} & -\dfrac{b'}{2} & 0 \end{bmatrix}. 
\end{equation}
To solve the continuity equations we diagonalize $A$ and find eigenvalues
\begin{equation} 
\lambda = 0, \pm \sqrt{ (-2a) \left[ \rho_H(1-\rho_H)b + \rho_L(1- \rho_L)b' \right] } .
\label{eqeigenVal}
\end{equation}
The eigenmodes or normal modes $\{ \phi_\alpha (x,t) \}$ satisfy traveling wave equations
\begin{equation*}
    \partial_t \phi_\alpha (x,t) = \lambda_\alpha \partial_x \phi_\alpha (x,t)
\end{equation*}
with $\alpha = 1,2,3$. Thus from linear hydrodynamics it follows that the time evolution of local density can be written as a linear superposition of three normal modes, also known as kinematic waves. One mode is stationary with propagation speed zero and two other modes move with equal and opposite velocities.

\begin{figure}[H]
\centering
\includegraphics[scale=1]{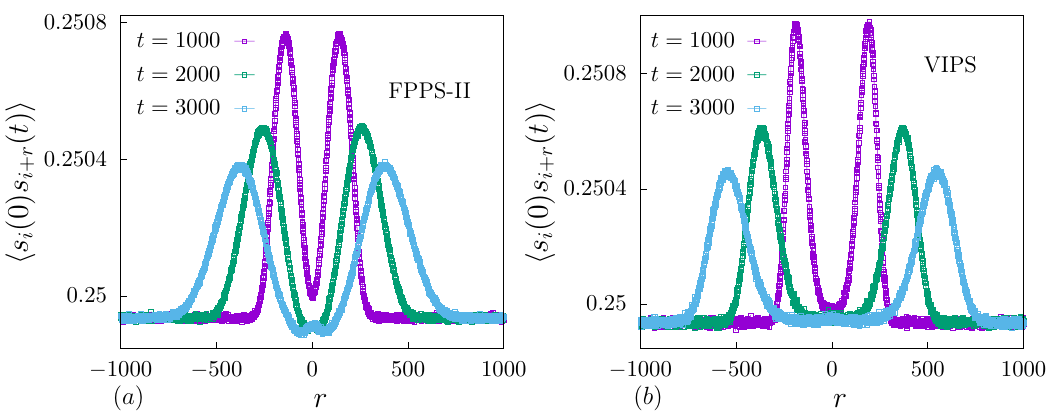}
\caption{Two-point unequal-time density correlation function of upslope bonds plotted against separation $r$. The splitting of the initial peak at $r = 0$ into two oppositely propagating waves of equal speed is taken as direct confirmation of two of the three kinematic-wave modes predicted by eq.\ref{eqeigenVal}. We used $b' = 0.3$ for FPPS-II (left), $b' = 0.18$ for VIPS (right) and $b = -0.4$ for both the cases, with  $N = 4104$.} \label{fig:momt}
\end{figure}

To check these results against simulations we show in Fig. \ref{fig:momt} the space time correlation of upslope bonds $\langle s_i(0) s_{i+r}(t) \rangle$ for few different $t$ during the coarsening phase. For $t=0$ this becomes an equal time correlation averaged over initial random configurations. A peak is expected at $r=0$ in this case. But as time goes on, our data in Fig. \ref{fig:momt}(a) and (b) show that for both FPPS-II and VIPS phase, the initial peak at origin splits into two peaks which move in opposite directions with equal speed. The third mode being stationary, can not be directly seen from these plots. However, the velocity of the peaks do not match exactly with the eigenvalues calculated for the current Jacobian matrix where mean field approximation was used. This is not surprising since long range order forms in the system as time goes on and mean field becomes more and more inaccurate. 

\newpage

\bibliographystyle{unsrt}

\end{document}